\documentclass[camera]{jpaper}

\usepackage[nocompress]{cite}
\usepackage{algorithmic}
\usepackage{array}

\makeatletter
\let\MYcaption\@makecaption
\makeatother

\usepackage[font=footnotesize]{subcaption}

\makeatletter
\let\@makecaption\MYcaption
\makeatother

\usepackage{fixltx2e}
\usepackage{dblfloatfix}
\usepackage[nolessnomore, italic]{mathastext}
\usepackage[T1]{fontenc}
\usepackage[usenames,dvipsnames,svgnames,table]{xcolor}
\usepackage{multirow}
\usepackage{hhline}
\usepackage[normalem]{ulem}
\usepackage{enumitem}
\usepackage[binary-units=true,parse-numbers=false]{siunitx}
\usepackage{setspace}
\usepackage{indentfirst}
\usepackage{footmisc}
\usepackage{fancyhdr}
\usepackage{authblk}
\usepackage[binary-units=true]{siunitx}
\usepackage[us,12hr]{datetime}
\usepackage[keeplastbox]{flushend}
\usepackage[hidelinks]{hyperref}

\newif\ifcameraready
\camerareadytrue
\newcommand{\versionnum}[0]{5}

\ifcameraready
\else
\fi

\ifcameraready
  \newcommand{\todo}[1][]{}
  \newcommand{\chIX}[1]{#1}
\else
  \newcommand{\todo}[1][]{\textbf{\fcolorbox{black}{red}{\color{white}{TODO}}} \underline{$\overline{\hbox{\emph{#1}}}$}}
  \newcommand{\chIX}[1]{{\color{BrickRed} #1}}
\fi

\newcommand{\chI}[1]{#1}

\newcommand{\chVIII}[1]{#1}

\ifcameraready

\else

\fi

\title{Experimental Characterization, Optimization, and Recovery\\of Data Retention Errors in MLC NAND Flash Memory
}

\date{}

\author{Yu Cai$^1$\qquad
Yixin Luo$^1$\qquad
Erich F. Haratsch$^2$\qquad
Ken Mai$^1$\qquad
Saugata Ghose$^1$\qquad
Onur Mutlu$^{3,1}$}
\affil{$^1$\emph{Carnegie Mellon University}\qquad
$^{2}$\emph{Seagate Technology}\qquad
$^{3}$\emph{ETH Z\"urich}}

\begin{document}

\maketitle

% \thispagestyle{fancy}
% \pagestyle{fancy}

%%%%%%%%%%%%%%%%%%%%%%%%%%%%%%%%%%%%%%%%%%%%%%%%%%%%%%%%%%%%%%%%%%%%%%%%%%%%%%%%
% sections
%%%%%%%%%%%%%%%%%%%%%%%%%%%%%%%%%%%%%%%%%%%%%%%%%%%%%%%%%%%%%%%%%%%%%%%%%%%%%%%%
% line-spacing

% !TEX root=paper.tex

\begin{abstract}

This paper summarizes our work on experimentally characterizing, mitigating,
and recovering data retention errors in multi-level cell (MLC) NAND flash
memory, which was published in HPCA 2015~\cite{cai.hpca15}, and examines the
work's significance and future potential.
Retention errors, caused by charge leakage over
time, are the dominant source of flash memory errors. Understanding,
characterizing, and reducing retention errors can significantly
improve NAND flash memory reliability and endurance.
In this work, we first characterize, with real 2Y-nm MLC
NAND flash chips, how the threshold voltage distribution of flash
memory changes with different retention \chVIII{ages} -- the length of time
since a flash cell was programmed. We observe from our characterization
results that 1)~the optimal read reference voltage of a
flash cell, using which the data can be read with the lowest raw
bit error rate (RBER), systematically changes with its retention
age, and 2)~different regions of flash memory can have different
retention ages, and hence different optimal read reference voltages.

Based on our findings, we propose two new techniques.
First, Retention Optimized Reading (ROR) adaptively learns and
applies the optimal read reference voltage for each flash memory
block online. The key idea of ROR is to periodically learn a tight
upper bound of the optimal read reference voltage, and from there approach the
optimal read reference
voltage. Our evaluations show that ROR can extend flash
memory lifetime by 64\% and reduce average error correction
latency by 10.1\%, with only 768~KB storage overhead in flash
memory for a 512~GB flash-based SSD. Second, Retention Failure
Recovery (RFR) recovers data with uncorrectable errors offline
by identifying and probabilistically correcting flash cells with
retention errors. Our evaluation shows that RFR reduces RBER
by 50\%, which essentially doubles the error correction capability,
and thus can effectively recover data from otherwise uncorrectable
flash errors.

\end{abstract}

%!TEX root=paper.tex
\section{Introduction}

Over the past decade, the capacity of NAND flash memory
has been increasing continuously, as a result of aggressive process
scaling and the advent of \emph{multi-level cell} (MLC) technology.
This trend has enabled NAND flash memory to replace
spinning disks for a wide range of applications -- from high
performance clusters and large-scale data centers to consumer
PCs, laptops, and mobile devices. Unfortunately, as flash density
increases, flash memory cells become more vulnerable to
various types of device and circuit level noise\chVIII{~\cite{R1,cai.date12, cai.procieee17, cai.procieee.arxiv17, cai.bookchapter.arxiv17}} -- e.g.,
retention noise~\cite{cai.date12,cai.iccd12,cai.itj13,mielke.irps08,papa.glsvlsi2014, cai.procieee17, cai.procieee.arxiv17, cai.bookchapter.arxiv17, luo.msst15}, read disturb noise~\cite{papa.glsvlsi2014, cai.dsn15, cai.hpca17, cai.procieee17, cai.procieee.arxiv17, cai.bookchapter.arxiv17}, cell-to-cell
program interference noise~\cite{cai.date12,cai.iccd13,cai.sigmetrics14, cai.hpca17, cai.procieee17, cai.procieee.arxiv17, cai.bookchapter.arxiv17}, and program/erase
(P/E) cycling noise~\cite{cai.date12,cai.date13, cai.procieee17, cai.procieee.arxiv17, cai.bookchapter.arxiv17}. These are sources of errors that can
significantly degrade NAND flash \chVIII{memory} reliability.

A traditional solution to overcome flash errors, regardless
of their source, is to use error-correcting codes (ECC)\chVIII{~\cite{R10,lin.book04, cai.procieee17, cai.procieee.arxiv17, cai.bookchapter.arxiv17}}.
By storing a certain amount of redundant bits per unit data,
ECC can detect and correct a limited number of raw bit errors.
With the help of ECC, flash memory can hide these errors from
the users until the number of errors per unit data exceeds the
correction capability of the ECC. Flash memory designers have
been relying on stronger ECC to compensate for lifetime reductions
due to technology scaling. However, stronger ECC,
which has higher capacity and implementation overhead, has
diminishing returns on the amount of flash lifetime improvement~\cite{
cai.iccd12,cai.itj13}. As such, we intend to look for more efficient ways
of reducing flash errors.

\begin{figure}[h]
  \centering
  \includegraphics[trim=0 400 300 0, clip, width=.9\columnwidth]{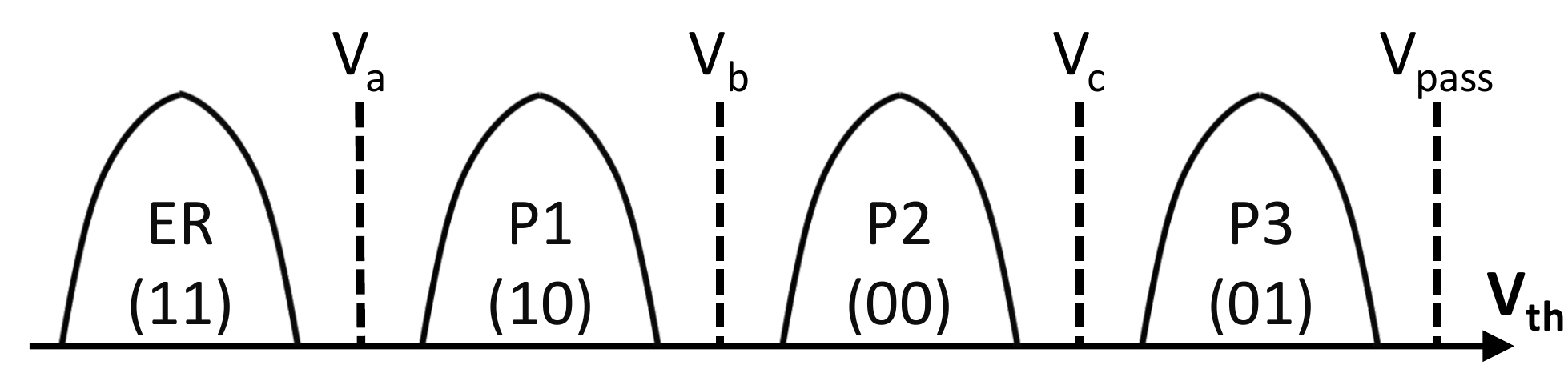}
  \caption{Threshold voltage distribution in 2-bit MLC NAND flash memory.  Stored data
  values are represented as the tuple (LSB, MSB). Reproduced from~\cite{cai.dsn15}.}
  \label{fig:vth}
\end{figure}

Retention errors, caused by charge leakage over time after a
flash cell is programmed, are the dominant source of flash
memory errors\chVIII{~\cite{cai.date12,cai.iccd12,cai.itj13,R12, cai.procieee17, cai.procieee.arxiv17, cai.bookchapter.arxiv17}}.
The amount of charge stored in a
flash memory cell determines the threshold voltage level of the
cell, which in turn represents \emph{the logical data value} stored in
the cell. As illustrated in Figure~\ref{fig:vth},
the threshold voltage ($V_{th}$) range of a 2-bit MLC NAND flash cell is divided
into four regions by three read reference voltages, $V_a$, $V_b$, and $V_c$. The
region in which the threshold voltage of a flash cell falls represents the
cell's current state, which can be ER (or erased), P1, P2, or P3.  Each state
decodes into a 2-bit value that is stored in the flash cell (e.g., 11, 10, 00,
or 01).\footnote{A detailed background on NAND flash memory design and operation,
and on data retention errors in NAND flash memory, can be found in our
prior works\chVIII{~\cite{cai.iccd12, cai.iccd13, cai.procieee17, cai.procieee.arxiv17, cai.bookchapter.arxiv17}}.}
% We represent this 2-bit value throughout the paper as a tuple (LSB,
% MSB), where LSB is the least significant bit and MSB is the most significant
% bit.  Note that the threshold voltage of all flash cells in a chip is bounded by
% an upper limit, $V_{pass}$, which is the \emph{pass-through voltage}.

As the manufacturing process technology for NAND flash memory scales
to smaller feature sizes, the capacitance of a flash cell, and the
number of electrons stored in the cell, decrease. State-of-the-art
MLC flash memory cells can \chVIII{store only} $\sim$100 electrons~\cite{cai.hpca15, min.fms14}. Gaining
or losing several electrons in a flash cell can significantly
change the cell's voltage level and eventually alter the state of
the cell. In addition, MLC technology reduces the size of the
\emph{threshold voltage window}~\cite{cai.date13}, i.e., the span of threshold voltage
values corresponding to each logical state, in order to store
more states in a single cell. This also makes the state of a cell
more likely to shift due to charge loss caused by retention
noise. As such, for NAND flash memory, retention errors are one of
the most important limiting factors of more aggressive process
scaling and MLC technology.

One way to reduce retention errors is to periodically read,
correct, and reprogram the flash memory before the number of
errors accumulated over time exceed the error correction capability
of the ECC, i.e., the maximum number of raw bit errors tolerable by the
ECC~\cite{cai.iccd12,cai.itj13,liu.fast12,pan.hpca12}. However, this \emph{flash
correct and refresh} (FCR) technique has two major limitations: 1)~FCR
uses a fixed read reference voltage to read data under different
retention ages, which is suboptimal, and
2)~FCR requires the flash controller to be consistently powered
on so that errors can be corrected, limiting its applicability to
enterprise deployments that have always-on power supplies.

In our HPCA 2015 paper~\cite{cai.hpca15}, we pursue a better understanding of retention
error behavior to improve NAND flash reliability and lifetime,
and find better \chVIII{(and complementary)} ways to mitigate flash retention errors. We
characterize 1)~the distortion of threshold voltage distribution
at different \emph{retention ages}, i.e., the idle time after the data is
programmed to the flash memory, for state-of-the-art 2Y-nm (20- to 24-nm)
NAND flash memory chips at room temperature, and 2)~the retention age
distribution of flash pages using disk traces
taken from real workloads. Our key findings are: 
\begin{enumerate}

\item Due to
threshold voltage distribution distortion, the \emph{optimal read reference
voltages} of flash cells, at which the minimum raw bit error rate (RBER) can be achieved, systematically shift to lower
values as retention age increases.

\item Pages within the same
flash block (the granularity at which flash memory can be
erased) tend to have similar retention ages and hence similar
optimal read reference voltages, whereas pages across different
flash blocks have different optimal read reference voltages.
\end{enumerate}

% The key ideas of our approach leverage these findings to 1)~optimize flash reliability, lifetime, and performance by learning
% and applying the optimal read reference voltage for each flash
% block online, and 2)~recover uncorrectable flash errors that
% exceed the correction capability of ECC by identifying and
% correcting fast- and slow-leaking cells offline (by comparing
% the distortion of threshold voltages of different flash cells over
% different retention ages).

Based on our findings, we propose two mechanisms to mitigate data
retention errors. First, we propose an \emph{online} technique called
\emph{Retention Optimized Reading} (ROR). They key idea of ROR is to reduce
the raw bit error rate by adaptively learning and applying the optimal read
reference voltage for each flash block. Our evaluations show that ROR
\chVIII{extends} flash lifetime by 64\% and \chVIII{reduces} average error correction latency by
10.1\%, with only 768~KB storage overhead for a 512~GB flash-based SSD\@.
Second, we propose an \emph{offline} error recovery technique called
\emph{Retention Failure Recovery} (RFR). The key idea of RFR is to identify
fast- and slow-leaking cells and \chVIII{probabilistically} determine the original value of an erroneous
cell based on its leakage-speed property and its threshold voltage. Our
evaluations show that RFR can effectively reduce the average raw bit error
rate (RBER) by 50\%, essentially doubling the error correction capability of
flash \chVIII{memory}, and allowing for the recovery of data otherwise uncorrectable by ECC\@.

\chVIII{We first summarize our experimental characterization results
(Section~\ref{sec:characterization}), and then introduce the Retention
Optimized Reading (Section~\ref{sec:ror}) and Retention Failure Recovery
(Section~\ref{sec:rfr}) techniques.}

\section{\chVIII{Flash Data} Retention Characterization}
\label{sec:characterization}

We use an FPGA-based flash memory testing platform to characterize real state-
of-the-art 2Y-nm NAND flash memory chips~\cite{cai.fccm11, cai.date12}. As absolute
threshold voltage values are proprietary information to NAND flash vendors, we
present our results using normalized voltages, where the nominal maximum value
of $V_{th}$ is equal to 512 in our normalized scale, and where 0 represents
GND\@.  Section~3.1 of our HPCA 2015 paper~\cite{cai.hpca15} provides a detailed
description of our experimental methodology.

\begin{figure}[h]
\centering
\includegraphics[trim=0 50mm 0 30mm,clip,width=0.48\textwidth]{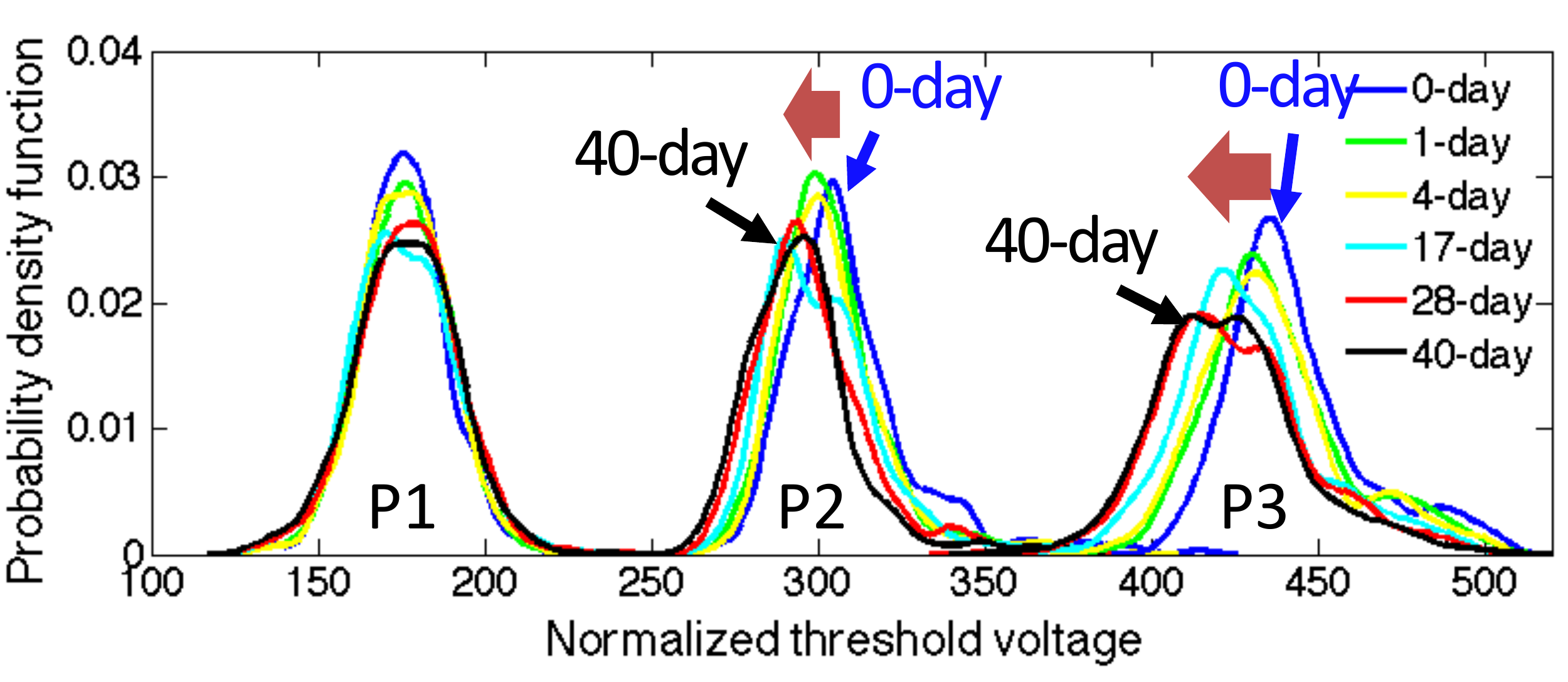}
\caption{
Threshold voltage distribution of 2Y-nm MLC NAND flash
memory vs.\ retention age, at 8K P/E cycles under room temperature.
Reproduced from~\cite{cai.hpca15}.
}
\label{fig:distribution}
\end{figure}

Figure~\ref{fig:distribution} shows the threshold voltage distribution of flash
memory at different retention ages for 8,000 P/E cycles.  
We make two observations from the figure. First, for the
higher-voltage states (P2 and P3), their threshold voltage distributions
systematically shift to lower voltage values as the retention age grows.  
Second, the distributions of each state become wider with higher retention
age, and that the distributions of states at higher voltage (e.g., P3) shift
faster than those of states at lower voltage (e.g., P1).

\begin{figure}[h]
\centering
\includegraphics[trim=0 120 20 0mm,clip,width=0.48\textwidth]{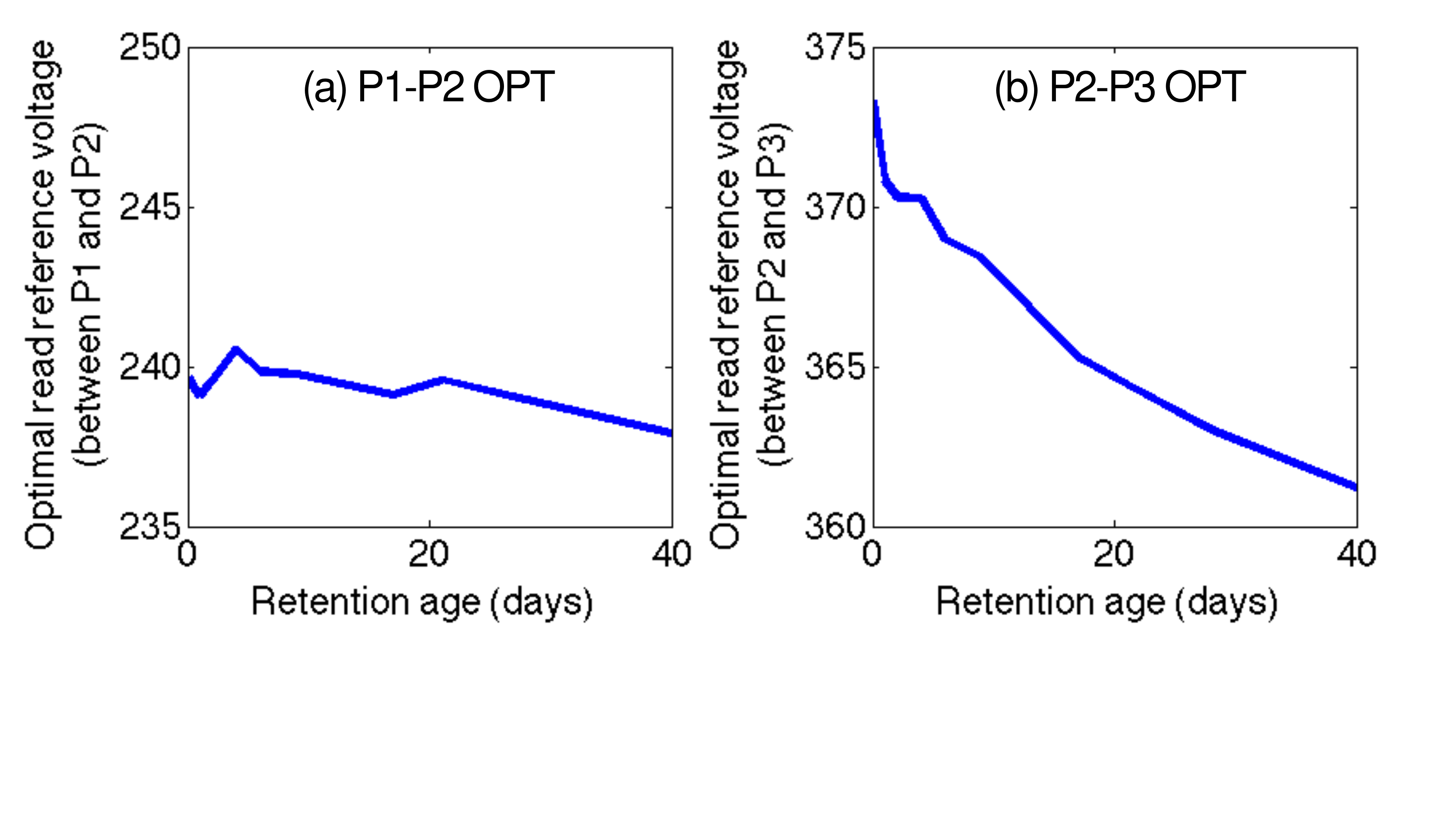}
\caption{
Effect of retention age on the optimal read reference voltage between
(a)~the P1 and P2 states, and (b)~the P2 and P3 states.
Reproduced from~\cite{cai.hpca15}.
}
\label{fig:fig5}
\end{figure}

We find that these changes due to retention leakage have an
impact to the \emph{optimal read reference voltage} (OPT),
which is the read reference voltage between two states that
minimizes the raw bit error rate (RBER). Figure~\ref{fig:fig5}
shows the optimal read reference voltage over retention age.
We make two observations from the figure.
First, Figure~\ref{fig:fig5}a shows a slightly decreasing trend of P1--P2 OPT (the
optimal read reference voltage used to distinguish between cells in the
P1 state and cells in the P2 state) over retention age. 
Second, we observe that P2--P3 OPT decreases much more rapidly with retention age
than P1--P2 OPT, as shown in
Figure~\ref{fig:fig5}b.

\begin{figure}[h]
\centering
\includegraphics[trim=0 360 250 0,clip,width=\linewidth]{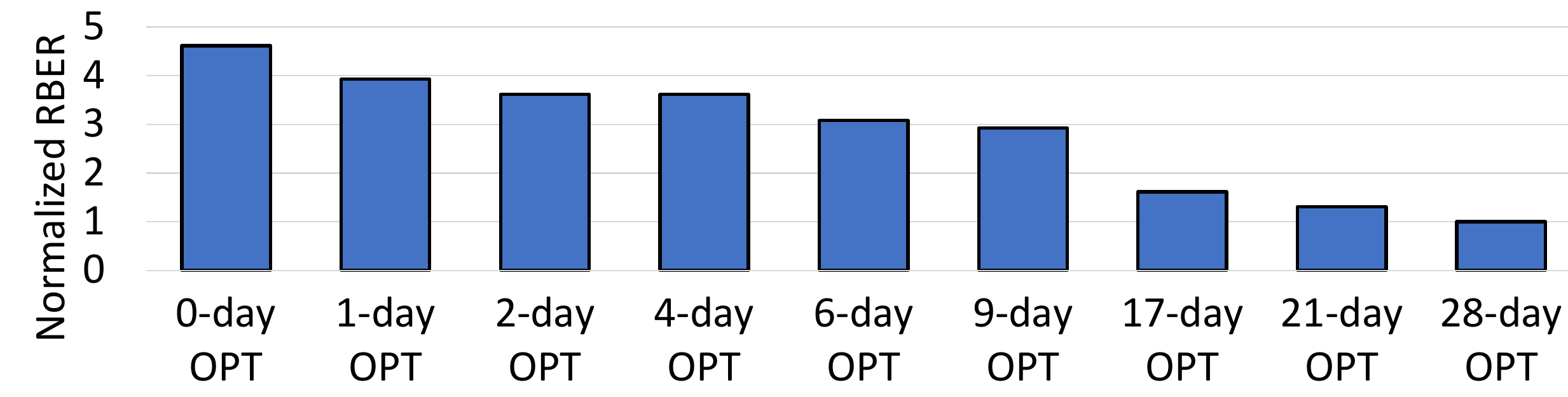}
\caption{Normalized RBER when reading 28-day-old data with different
optimal read reference voltages (normalized to 28-day OPT).
Reproduced from~\cite{cai.hpca15}.}
\label{fig:fig6}
\end{figure}

As the distributions continue to shift with growing retention age, the OPT for
one retention age will be different than the OPT for a different age,
suggesting that a dynamically changing OPT is ideal. To quantify how the
choice of read reference voltage affects RBER, we apply the optimal read
reference voltages (OPTs) determined for \{0, 1, 2, 6, 9, 17, 21, 28\}-day
retention ages to read 28-day-old data. Figure~\ref{fig:fig6} shows the RBER
obtained when reading the 28-day-old data with different OPTs, normalized to
the RBER obtained when reading the data with the 28-day OPT\@. This figure
shows that picking the correct value of OPT for each retention age results in
a lower RBER. In turn, this allows us to \emph{extend} the
lifetime (i.e., the number of P/E cycles the device can tolerate) of the NAND flash
memory if we always use
the correct OPT based on the retention age of the data that is being read.

In Section~3 of our HPCA 2015 paper~\cite{cai.hpca15}, 
we perform several other \chVIII{experimental characterization studies of flash
memory data} retention behavior, and make the
following eight \chVIII{new} findings:

\begin{enumerate}[leftmargin=1.2em]
  
   \item The threshold voltage distributions of the P2 and P3 states
     systematically shift to lower voltages with retention age.  
  
   \item The threshold voltage distribution of each state becomes wider with
    higher retention age.

  \item The threshold voltage distribution of a higher-voltage state shifts
    faster than that of a lower-voltage state.

   \item Both P1--P2 OPT and P2--P3 OPT become smaller over retention age.

   \item P2--P3 OPT changes more significantly over retention age than P1--P2 OPT\@.

  \item The optimal read reference voltage corresponding to one retention age is
    suboptimal (i.e., it results in a higher RBER) for reading data with a
    different retention age.

  \item RBER becomes lower when the retention age for which the used read
    reference voltage is optimized becomes closer to the actual retention age of
    the data.

   \item The lifetime of NAND flash memory can be extended if the optimal
     read reference voltage that corresponds to the retention age of the data is
     used.

\end{enumerate}

\section{Retention Optimized Reading (ROR)}
\label{sec:ror}

To optimize flash \chVIII{memory} performance without compromising flash lifetime, we first
breakdown and analyze the components of the flash \chVIII{memory} read latency. 
A read operation typically makes use of the {read-retry}
operation\chVIII{~\cite{cai.date13, fukami2017improving, cai.procieee17,
cai.procieee.arxiv17, cai.bookchapter.arxiv17}}, which
performs multiple data read attempts using different read reference voltages
until the read succeeds (i.e., ECC successfully corrects all of the raw bit 
errors).
\chVIII{A} detailed analysis of the flash \chVIII{memory} read latency
can be found in Section~4.1 of our HPCA 2015 paper~\cite{cai.hpca15}. We summarize the following four
observations from this analysis:

\begin{itemize}[leftmargin=1.2em] 
  \setlength\itemsep{0pt}

  \item The read latency of NAND flash \chVIII{memory} can be reduced by minimizing the number of reads
  performed during read-retry.

  \item The number of reads can be reduced by using a closer-to-optimal
    starting read reference voltage \chVIII{in the read-retry process}.

  \item The optimal read reference voltages of pages in the same block are
    close, while those of pages in \chVIII{\emph{different}} blocks are \chVIII{\emph{not}} always close.

  \item The optimal read reference voltage of pages in a block is upper-bounded
    by the optimal read reference voltage of the page in the block that was programmed \chVIII{\emph{last}}.

\end{itemize}

Based on these observations, we propose \emph{Retention Optimized Reading}
(ROR), which consists of two components: 1)~an online pre-optimization
algorithm that learns the \chVIII{\emph{starting}} read
reference voltage for each block, and 2)~an improved read-retry technique that
uses the starting read reference voltage to reduce the search space of OPT
\chVIII{(i.e., the optimal read reference voltage)} for
the block.
Section~4.2 of our HPCA 2015 paper~\cite{cai.hpca15} provides a detailed description of the
components of ROR.  We briefly summarize the components below.

The first component, the online pre-optimization algorithm, is triggered both daily
and after power-on for each block. This algorithm consists of the following
four steps:

\begin{itemize}[leftmargin=1.2em]

\item \emph{Step~1:} The flash controller first reads 
the highest-numbered page in a flash block (e.g., page~255 in a block that contains 256 pages), with
any default read reference voltage $V_{default}$, and attempts to correct the
errors in the raw data read from the page. We chose the highest-numbered page in the block because it is
programmed last, and, thus, has the \chVIII{lowest} retention age \chVIII{and the
highest OPT value within the block. Hence, we use the OPT for the highest-numbered
page as a tight \emph{upper bound} of OPT for the block.} Next, we record the
number of raw bit errors as the current lowest error count ($N_{ERR}$), and
the applied read reference voltage as $V_{ref}$ = $V_{default}$. If we cannot
find the error count (i.e., the error is uncorrectable), we record the maximum
number of errors correctable by ECC as $N_{ERR}$.

\item \emph{Step~2:} The controller tries to read the page using a lower read reference voltage.
Since
we want to find the optimal read reference voltage for the highest-numbered page in the block, we approach it
from the current starting read reference voltage step by step. Since OPT
typically decreases over retention age, we first attempt to lower the read
reference voltage. We decrease the read reference voltage
to ($V_{ref} - \Delta V$) and read the highest-numbered page. If the number of corrected
errors in the new data is less than or equal to the old $N_{ERR}$, we
update $N_{ERR}$ and $V_{ref}$ with the new values. We repeat Step~2
until the number of corrected errors in the new data is greater
than the previous value of $N_{ERR}$, or the lowest possible read reference voltage
is reached.

\item \emph{Step~3:} The controller tries to read the page using a higher read reference voltage. Since
the optimal threshold voltage might increase in rare cases, we
also attempt to increase the read reference voltage. We increase
the read reference voltage to ($V_{ref} + \Delta V$) and read the highest-numbered page in the block.
Again, if the number of corrected errors in the new data is less
than or equal to $N_{ERR}$, we update $N_{ERR}$ and $V_{ref}$ with the new
values. We repeat Step~3 until the number of corrected errors in
the new data is greater than the previous value of $N_{ERR}$, or the highest possible
read reference voltage is reached.

\item \emph{Step~4:} Record the optimal read reference voltage. After
Step~3, the most recently-used value of $V_{ref}$ is the optimal read reference
voltage for the highest-numbered page. Thus, we record this voltage as the \chVIII{\emph{upper
bound}} of the optimal read reference voltages for the block.

\end{itemize}

The second component is an improved read-retry technique that takes
advantage of the recorded starting read reference voltage. During a
normal read operation, the flash controller first attempts to read the
data with the recorded starting read reference voltage. Then,
since the recorded starting read reference voltage is the upper
bound of the OPTs within the block, we iteratively decrease the
read reference voltage until the read operation succeeds. Note
that the starting read reference voltages are accessed frequently
(on each read operation) by the flash controller, so we store
them in the SSD's DRAM buffer to allow fast access.

% ROR reduces the read latency of NAND flash memory.

% \section{Results}

Our key evaluation results show that ROR achieves the same flash lifetime
improvements as naive read-retry, which \chVIII{has a read latency that is} 64\% longer than a baseline that
uses a fixed read reference voltage. Due to a reduction in raw bit error rate,
ROR reduces the ECC decoding latency by 10.1\% on average compared to the
baseline, which is equivalent to a 2.4\% reduction in overall flash read
latency. Compared with the original read-retry technique, which we explain in detail
in Section~4.1 of our HPCA 2015 paper~\cite{cai.hpca15}, ROR reduces the read-retry operation
count by 70.4\%, and thus reduces the overall read latency by the same
fraction.
This reduction is due to two reasons: 1)~ROR
starts the read-retry process at a close-to-optimal starting read
reference voltage that is estimated and recorded daily and upon power-on; and
2)~ROR approaches OPT in a known, informed direction from this starting read
reference voltage.

Section~4.4 of our HPCA 2015 paper~\cite{cai.hpca15} provides more
results from our evaluation of ROR.
In our HPCA 2015 paper, we show that the performance overhead of ROR,
which is periodically triggered by an online pre-optimization algorithm, can
\chVIII{be largely}
hidden by executing the algorithm only when the SSD is idle, \chVIII{or
in} the background
at a lower priority. This is because, even considering the worst-case scenario,
we obtain an estimated pre-optimization latency of 3, 15, and 23
seconds for flash memory with a 1-day, 7-day, and 30-day-equivalent retention
age, respectively.
Since the flash pages within a block is programmed at similar times, the
optimal read reference voltages of these pages are close. So we store one
byte per block for each starting read reference voltage
learned for the ER-P1 OPT, the P1--P2 OPT, and the P2--P3 OPT.
We also show that ROR requires only 768~\texttt{KB} of storage overhead, to
store the entire read reference voltage table for an assumed 512~\texttt{GB}
flash drive.

\section{Retention Failure Recovery (RFR)}
\label{sec:rfr}

Even with ROR, the retention error rate will eventually exceed the ECC limit
as retention age keeps increasing. At that
point, some reads will have more raw \chVIII{errors than} can be corrected
by ECC, preventing the drive from returning the data to the user.
Traditionally, this would be the point of \emph{data loss} \chVIII{and thus
the end of flash memory lifetime}.

\chVIII{We show that retention failure is
avoidable under various circumstances. 
In Section~5.1 of our HPCA 2015 paper~\cite{cai.hpca15}, we show that high
temperature can significantly increase the number of retention errors in a
short period of time, which leads to unexpected data loss. For example, if the
required refresh period of the flash memory is one week at room temperature,
uncorrectable errors may start to accumulate after a mere 36~minutes. We also
discuss why completely avoiding such retention failure is
unrealistic.
\chVIII{No} previous technique can prevent
data loss \chIX{\emph{after}} retention failure happens.

We introduce \emph{Retention Failure Recovery} (RFR), which
\emph{enables} us to recover data from a failed flash page \emph{offline} after the number of
errors in the page exceed the total number of errors that ECC can correct.
Due to process variation, different
flash cells on the same chip can have different charge leakage speeds.
We describe a technique to classify
fast- and slow-leaking cells in just a few days, which enables RFR to \chVIII{probabilistically \emph{infer}} the
original value stored in each flash cell. Our evaluation, based on data from real NAND
flash chips, shows that RFR can reduce raw bit error rate by 50\%, and thus ECC
can then be used to recover a majority of the data \chVIII{in pages with retention failures}.}

\begin{figure}[h]
  \centering
%   \vspace{-5pt}%
  \includegraphics[width=\linewidth]{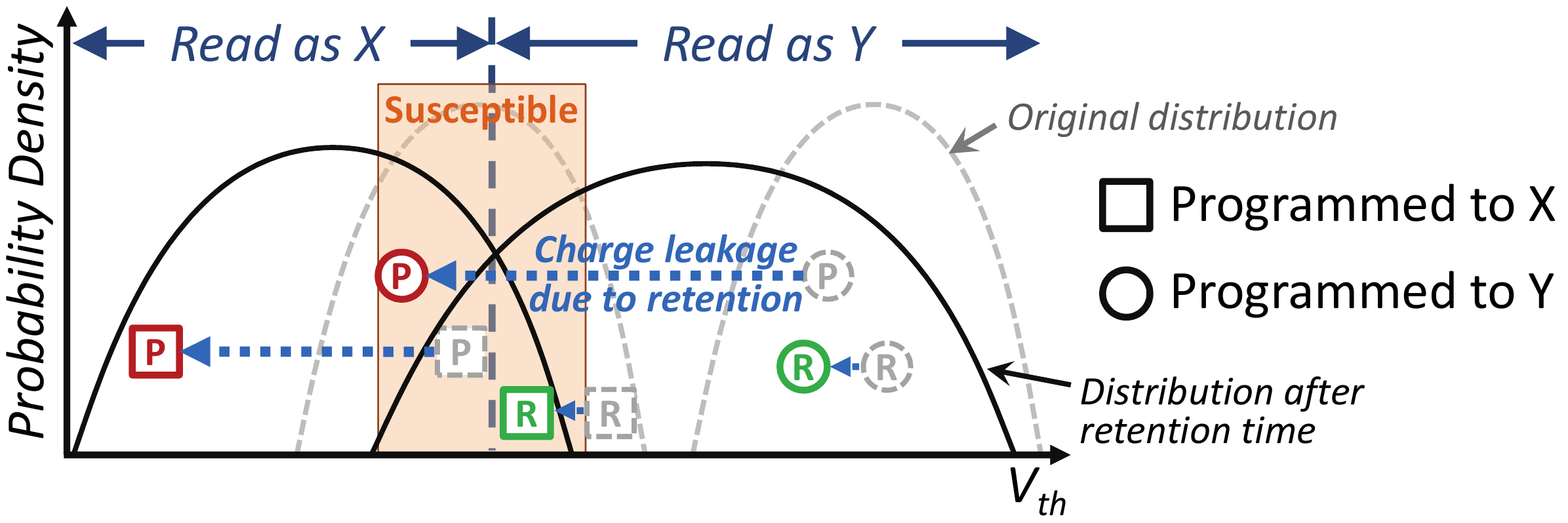}%
  % \vspace{-5pt}%
  \caption{Some retention-prone (P) and retention-resistant (R) cells
are incorrectly read after charge leakage due to retention time.
RFR identifies and corrects the incorrectly read cells based on their
leakage behavior. Reproduced from~\cite{cai.procieee17}.}%
  \label{fig:F33}%
%  \vspace{-5pt}%
\end{figure}

Figure~\ref{fig:F33} shows how the threshold voltage of a retention-prone
cell (i.e., a \emph{fast-leaking} cell, labeled P in the figure)
decreases over time (i.e., the cell shifts to the left) due to
retention leakage, while the threshold voltage of a retention-
resistant cell (i.e., a \emph{slow-leaking} cell, labeled R in the
figure) does \chVIII{\emph{not}} change significantly over time. Retention
Failure Recovery (RFR) uses this classification of retention-prone
versus retention-resistant cells to correct the
data from the failed page \emph{without} the assistance of ECC.
Without loss of generality, let us assume that we are studying
susceptible cells near the intersection of two threshold
voltage distributions X and Y, where Y contains higher voltages
than X. Figure~\ref{fig:F33} highlights the region of cells considered
susceptible by RFR using a box, labeled \emph{Susceptible}.
A susceptible cell within the box that is retention prone
likely belongs to distribution Y, as a retention-prone cell
shifts rapidly to a lower voltage (see the circled cell labeled
P within the \emph{susceptible} region in the figure). A retention-resistant
cell in the same \emph{susceptible} region likely belongs
to distribution X (see the boxed cell labeled R within the
\emph{susceptible} region in the figure).

\begin{figure}[h]
\centering
\includegraphics[trim=20mm 65mm 0 0,clip,width=\linewidth]{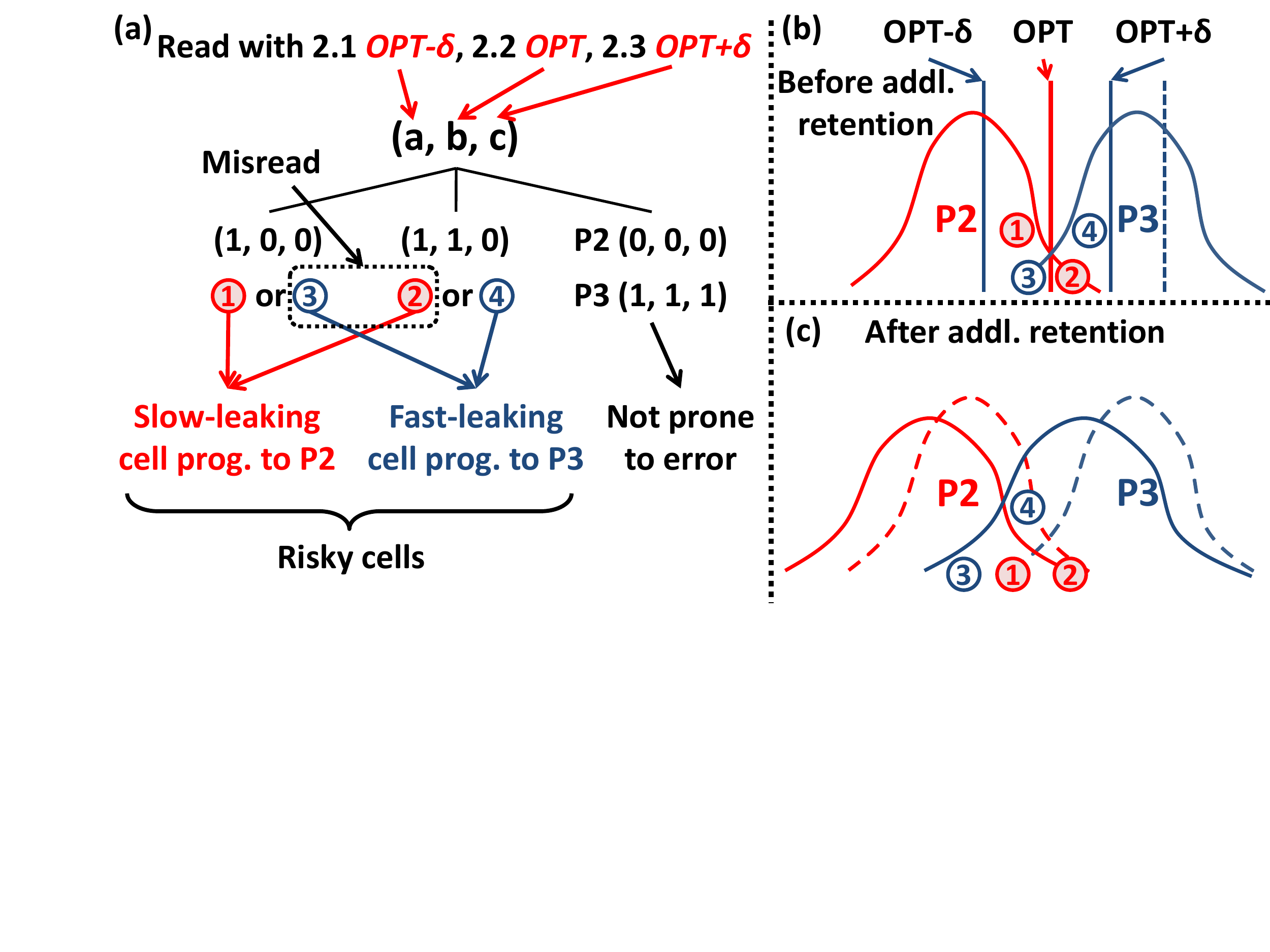}
\caption{\chVIII{(a)~Classification of risky (i.e., susceptible) cells to identify misread bits, 
(b)~cells before additional retention loss, and (c) cells after additional retention loss.}
Reproduced from~\cite{cai.hpca15}.}
\label{fig:RFR}
\end{figure}

RFR identifies fast- vs.\ slow-leaking cells, and uses selective
bit flipping to correct retention failures, thus reducing RBER\@. With reduced
raw bit errors, the read data may be reconstructed by ECC with a higher
probability. RFR consists of the following four offline steps, which are triggered
when an uncorrectable error is found:

\begin{itemize}[leftmargin=1.2em]
  \setlength\itemsep{0pt}

  \item \emph{Step~1:} Identify data with a retention failure. Once \chVIII{the flash}
    controller fails to read a flash page, a retention failure is identified on
    that page.

  \item \emph{Step~2:} Identify \chVIII{susceptible} \chIX{cells}
    % risky cells \chVIII{(i.e., cells with potential
    % errors due to retention loss)} 
    using three read operations. We read the
    \chVIII{failed page using three read reference voltages: OPT (the optimal
    read reference voltage) minus some margin $\delta$ (Step~2.1),
    OPT (Step~2.2),
    and OPT plus $\delta$ (Step~2.3). 
    The value of $\delta$ is large enough to include the entire \emph{Susceptible}
    region shown in Figure~\ref{fig:F33}.}
    % failed page with read reference voltage set to optimal read reference
    % voltage (OPT, \chVIII{Step~2.2}) plus \chVIII{(Step~2.3)} and minus
    % \chVIII{(Step~2.1)} some margin to identify those risky cells that
    % are more likely to contain errors. 
    Figure~\ref{fig:RFR}a illustrates the
    identification of \chVIII{susceptible (i.e., risky)} cells, which are denoted as type \textcircled{1},
    type \textcircled{2}, type \textcircled{3}, and type \textcircled{4} cells.

  \item \emph{Step~3:} Identify fast- and slow-leaking cells. We compare the threshold
    voltage of \chVIII{susceptible} cells before and after several days of retention to
    classify them as fast- and slow-leaking cells. Figures~\ref{fig:RFR}b
    and \ref{fig:RFR}c illustrate how \chIX{the cells} shift differently after additional
    retention loss. Among the \chVIII{susceptible} cells, type \textcircled{1} and type
    \textcircled{2} cells are slow-leaking cells, whereas type \textcircled{3}
    and type \textcircled{4} cells are fast-leaking cells.

  \item \emph{Step~4:} Selectively flip bits based on the identification \chVIII{results} from Step~3. Using the leakage speed information, we
    now know that type \textcircled{2} and type \textcircled{3} cells are likely
    misread. Thus\chVIII{,} we simply flip those cells to correct these likely errors.

\end{itemize}

\begin{figure}[h]
\centering
\includegraphics[trim=0 280 220 0,clip,width=\linewidth]{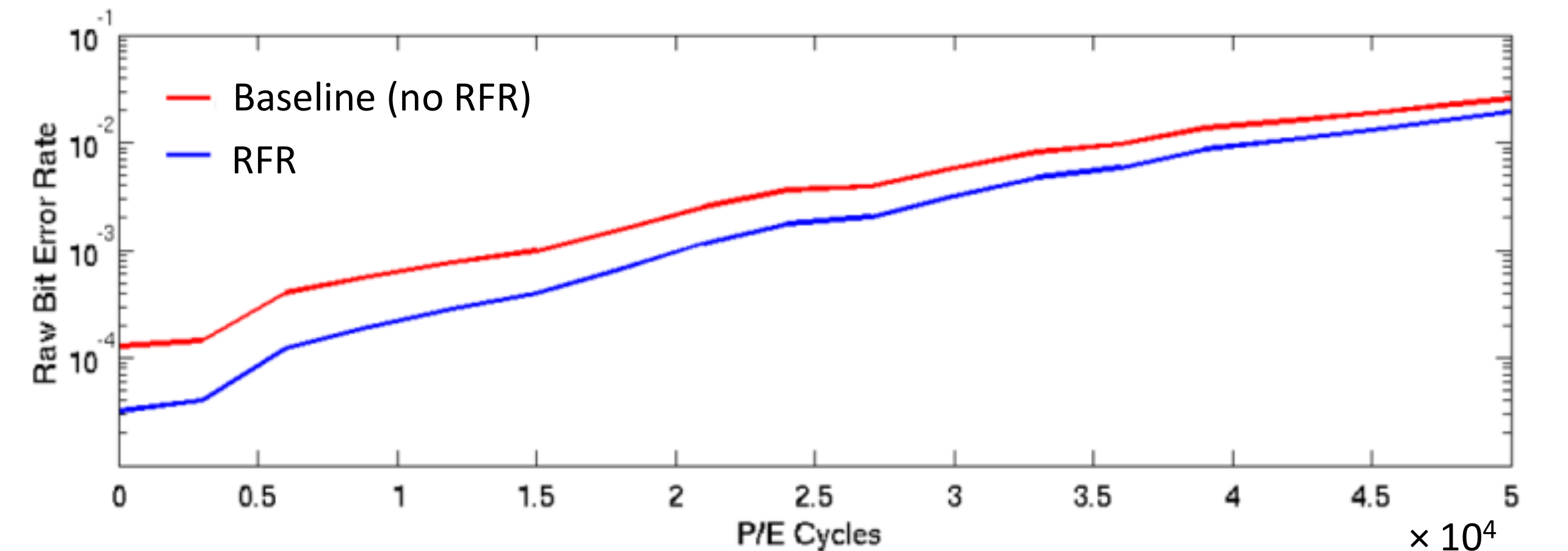}
\caption{
Effect of the RFR technique on raw bit error rate.
Reproduced from~\cite{cai.hpca15}.
}
\label{fig:fig16}
\end{figure}

We evaluate RFR on data programmed to random values that has 28-day equivalent
retention age. In Step 3, we introduce an additional 12 days' worth of
equivalent retention age. Figure~\ref{fig:fig16} shows the resulting raw bit
error rate of RFR over a range of P/E cycles (compared to that of the
baseline). This figure shows that RFR reduces the RBER by 50\%, averaged
across \chVIII{all evaluated wearout} levels (P/E cycles). Thus, we expect the number of raw
bit errors to be halved, increasing the chances that these errors are
correctable by ECC\@.

%!TEX root=paper.tex

\section{Related Work}

To our knowledge, our HPCA 2015 paper~\cite{cai.hpca15} 
\chVIII{is} the first to
\chVIII{1)}~experimentally characterize and comprehensively analyze how the threshold
voltage distribution changes \emph{over different retention ages}, as well as
the implication of these changes on the read reference voltage and lifetime,
using real state-of-the-art 2Y-nm MLC NAND flash \chVIII{memory chips}; and
2)~proposes two novel techniques to mitigate the impact of retention age online
and to recover from data loss by exploiting retention behavior.
In this section, we briefly discuss \chVIII{various} related works.
% We categorize related work into six major categories: (1)~retention error
% characterization, (2)~NAND flash memory error characterization, (3)~3D NAND
% error characterization, (4)~retention error mitigation using
% periodic refresh, (5)~read reference voltage optimization, and
% (6)~error recovery.

\subsection{Works on NAND Flash Memory}

\textbf{\chVIII{NAND Flash Memory} Retention Error Characterization.} \chVIII{Multiple} prior works
characterize NAND flash data retention, but \chVIII{mainly} in terms of
RBER~\cite{cai.date12, cai.iccd12, cai.itj13, mielke.irps08}. These works show
that 1)~retention errors are the dominant errors in NAND flash memory, and
2)~the retention error rate increases with the retention age and the P/E
cycle. Papandreou et al.~\cite{papa.glsvlsi2014} characterize the retention effect on threshold
voltage distributions under high temperature baking,
and find that the distribution shifts to lower voltage over retention time,
and so does the optimal read reference voltage. In contrast, our HPCA 2015 paper~\cite{cai.hpca15}
characterizes data retention under room temperature, which is closer to how
NAND flash memories are typically used~\cite{cai.hpca15}. Our recent work
characterizes how data retention affects the threshold voltage distribution
for TLC NAND flash memory~\cite{cai.procieee17, cai.procieee.arxiv17, cai.bookchapter.arxiv17}, making similar findings as
our HPCA 2015 paper~\cite{cai.hpca15}.

\textbf{NAND Flash Memory Error Characterization.} Prior \chVIII{works
study} different types of NAND flash memory errors in MLC, planar NAND flash
memory, including P/E cycling errors~\cite{mielke.irps08, cai.date13,
parnell.globecom14, luo.jsac16, papa.glsvlsi2014}, programming
errors~\cite{cai.hpca17, parnell.globecom14, luo.jsac16}, cell-to-cell program
interference errors~\cite{cai.iccd13, cai.date13, cai.sigmetrics14}, retention
errors~\cite{mielke.irps08, cai.date13, cai.iccd12, cai.hpca15,
papa.glsvlsi2014}, and read disturb errors~\cite{mielke.irps08, cai.dsn15,
papa.glsvlsi2014}. These works characterize how raw bit error rate and
threshold voltage \chVIII{distributions} change \chVIII{with} various types of noise. Our recent work
characterizes the same types of errors in \chVIII{planar TLC} NAND flash memory and
has similar findings~\cite{cai.procieee17, cai.procieee.arxiv17, cai.bookchapter.arxiv17}. Thus, we believe that most of the
findings on MLC NAND flash memory can be generalized to any types of planar
NAND flash memory devices (e.g., SLC, MLC, TLC, or QLC).
Recent works~\cite{meza2015large, narayanan2016ssd, schroeder2017reliability} have also studied SSD errors in the field, and have shown the
system-level implications of these errors \chVIII{in} large-scale
data centers.
\chVIII{Unlike our characterization, these in-the-field studies do \chIX{\emph{not}} have 
access to the underlying NAND flash memory
within the SSDs that they test, and, thus, are unable to show detailed data retention behavior.}

\textbf{3D NAND \chVIII{Flash Memory} Error Characterization.} Recently, manufacturers
have begun to produce SSDs that contain \emph{three-dimensional} (3D) NAND
flash memory~\cite{yoon.fms15, park.jssc15, kang.isscc16, im.isscc15,
micheloni.procieee17, micheloni.sn16}.  In 3D NAND flash memory,
\emph{multiple layers} of flash cells are stacked vertically to increase the
density and to improve the scalability of the memory~\cite{yoon.fms15}.  In
order to achieve this stacking, manufacturers have changed a number of
underlying properties of the flash memory design. We refer \chVIII{readers} to our
prior work for a \chVIII{detailed} comparison between 3D NAND flash memory and planar
NAND flash memory~\cite{cai.procieee17, cai.procieee.arxiv17, cai.bookchapter.arxiv17}. 
Previous works~\cite{mizoguchi.imw17, choi.vlsit16} compare the retention loss between 3D
charge trap NAND flash memory and planar NAND flash memory through real device
characterization, and find that 3D charge trap
cells leak charge faster than planar NAND cells and thus \chVIII{experience the phenomenon of} \emph{early
retention loss}. Our recent work\chIX{~\cite{luo.hpca18}} characterizes the impact of dwell time, i.e., the
idle time between consecutive program cycles, and environmental temperature on
the retention loss speed and program \chVIII{variation} of 3D charge trap NAND flash
memory, \chVIII{and proposes techniques to mitigate these issues to improve
flash memory lifetime}.
Recent work~\cite{xiong.sigmetrics17} characterizes the latency and raw bit error
rate of 3D NAND flash memory devices based on floating gate cells,
and makes similar observations as those for planar NAND flash memory devices based on floating
gate cells.
Prior works have reported several differences between 3D NAND and planar NAND
through circuit level measurements, including \chVIII{the fact that 3D NAND
flash cells exhibit} 1)~smaller
program variation at high P/E cycle~\cite{park.jssc15}, 2)~smaller program
interference~\cite{park.jssc15}, and 3)~early retention loss~\cite{choi.vlsit16, mizoguchi.imw17, choi.vlsit16}.
The field (both academia and industry) is currently in much need of \chVIII{detailed} rigorous
experimental characterization and analysis of \chVIII{state-of-the-art} 3D NAND flash memory devices.

\textbf{Retention Error Mitigation Using Periodic
Refresh.} Prior works~\cite{cai.iccd12,cai.itj13,liu.fast12,pan.hpca12}
propose to use periodic refresh to mitigate retention errors. \chVIII{Cai et
al.~\cite{cai.iccd12, cai.itj13} introduce} 1)~\emph
{remapping-based refresh}, which periodically reads data from each valid flash
block, corrects any data errors, and \emph{remaps} the data to a different
physical location, 2)~\emph{in-place refresh}, which incrementally replenishes
the lost charge of each page at its current location, and 3)~\emph{adaptive
refresh}, which allows the controller to adaptively adjust the rate that the
refresh mechanisms are invoked based on the wearout (i.e., the current P/E
cycle count) of the NAND flash memory~\cite{cai.iccd12,cai.itj13}; or the
temperature of the SSD~\cite{cai.date12,cai.hpca15}. However, these techniques
1)~require the system to be consistently powered on, and 2)~are unaware of the
fact that the optimal read reference voltage changes with different retention
age. Note that these works always apply a \chVIII{\emph{fixed}} read reference voltage
regardless of the retention age of the cell, which is suboptimal for reading
flash blocks at \chVIII{\emph{different}} retention ages. In contrast, our ROR technique
optimizes the read reference voltage of each flash block based on its
retention age, leading to significant lifetime improvements.
Several works~\cite{luo.msst15, choi2017exploiting, shi2016retention} find that refresh operations consume a large number of P/E cycles,
and propose techniques that exploit workload write-hotness to relax the
guaranteed retention time of NAND flash memory without requiring
refresh.
For example, WARM~\cite{luo.msst15} partitions write-hot and write-cold data using a lightweight
mechanism designed for flash memory, and \chVIII{\emph{eliminates}} the need to refresh
write-hot data, leading to significant lifetime improvements over existing
periodic refresh mechanisms. \chVIII{Our techniques can be combined with such
refresh elimination techniques for higher lifetime and performance.}

\textbf{Read Reference Voltage Optimization.}
A few works~\cite{papa.glsvlsi2014,cai.iccd13,cai.sigmetrics14} propose
optimizing the read reference
voltage.
Cai et al.~\cite{cai.sigmetrics14} propose a technique to calculate the optimal
read reference voltage from the mean and variance of the
threshold voltage distributions, which are characterized by the
read-retry technique~\cite{cai.date13}. The cost of such a technique is relatively
high, as it requires periodically reading flash memory
with all possible read reference voltages to discover the threshold
voltage distributions. Papandreou et al.~\cite{papa.glsvlsi2014} propose to apply
a per-block close-to-optimal read reference voltage by periodically
sampling and averaging 6 OPTs within each block,
learned by exhaustively trying all possible read reference voltages.
In contrast, ROR can find the actual optimal read reference
voltage at a much lower latency, thanks to the new findings and observations in
our HPCA 2015 paper~\cite{cai.hpca15}. We show that ROR greatly
outperforms naive read-retry. \chVIII{The latter is}
significantly simpler than the mechanism proposed in \cite{papa.glsvlsi2014}.

Recently, Luo et al.~\cite{luo.jsac16} propose to accurately predict the optimal read
reference
voltage using an online flash channel model for each chip learned
online. Cai et al.~\cite{cai.dsn15} propose a new technique called
$V_{pass}$ tuning, which tunes
the \emph{pass-through voltage}, i.e., a high reference voltage applied to
turn on unread cells in a block, to mitigate read disturb
errors.
Du et al.~\cite{du2017reducing} propose to tune the optimal read reference voltages for ECC
soft decoding to improve the ECC correction capability (i.e., the maximum number of errors
that ECC can correct).
Fukami et al.~\cite{fukami2017improving} propose to use read-retry to improve the
reliability of the chip-off forensic analysis of NAND flash memory
devices. \chVIII{Our proposals are complementary to all these techniques.}

\textbf{Error Recovery.} 
% No mechanism prior to
% our HPCA 2015 paper (e.g.,~\cite{cai.iccd12, cai.itj13, papa.glsvlsi2014,
% liu.fast12, pan.hpca12, cai.dsn15}) can recover the data from an uncorrectable
% error that is beyond the error correction capability of ECC caused by data
% retention. 
To our knowledge, our HPCA 2015 paper~\cite{cai.hpca15} proposes the first
mechanism that can recover data even \chVIII{\emph{after}} ECC is unable to successfully correct
all of the errors due to retention \chVIII{loss}.
One of our works~\cite{cai.dsn15} \chVIII{builds on our HPCA 2015 paper and} adapts the RFR mechanism to
opportunistically recover from \emph{read disturb errors} instead of
retention errors. FlashDefibrillator (FD)~\cite{jeong2015flashdefibrillator} improves upon RFR to recover
from data retention errors \emph{online}.
FD recovers data retention errors online by applying a sequence of diagnostic
pulses that recharge the fast-leaking cells. This helps recover \chVIII{otherwise} uncorrectable
errors in two ways: (1)~fast-leaking cells may be recharged back to the correct
state, (2)~fast-leaking cells recharge faster than slow-leaking cells, thus
fast-leaking cells can be identified as the cells whose threshold \chVIII{voltages}
increase faster during the diagnostic pulses. \chVIII{These two more recent
works~\cite{cai.dsn15,jeong2015flashdefibrillator} directly build upon our HPCA 2015 paper.}

% However, we observe that such uncorrectable errors can happen
% quickly under unexpected high temperature, as the effective retention age can
% easily become longer than the refresh period or the assumed retention target.

% \noindent \textbf{Other.} The discussion of read-retry technique in our HPCA
% paper has inspired recent work to use read-retry to improve the
% reliability of chip-off forensic analysis of NAND flash memory
% devices~\cite{fukami2017improving}. The discussion of ECC soft decoding in SSD
% controller in our HPCA paper has inspired recent work to improve NAND
% flash memory read performance by reducing retention error rate.

\subsection{Data Retention Errors in DRAM}
\label{sec:othermem:retention}

DRAM uses the charge
within a capacitor to represent one bit of data. Much like the
floating gate within NAND flash memory, charge leaks from
the DRAM capacitor over time, leading to data retention
issues. Unlike a NAND flash cell, where leakage typically
leads to data loss after several days to years of retention
time, leakage from a DRAM cell leads to
data loss after a retention time on the order of \emph{milliseconds} to
\emph{seconds}~\cite{liu.isca13}.

\chI{The retention time of a DRAM cell depends upon several factors\chVIII{~\cite{liu.isca13}},
including (1)~manufacturing process variation and 
(2)~temperature.
Manufacturing process variation affects the amount of current that leaks
from each DRAM cell's capacitor and access transistor~\cite{liu.isca13}.
As a result, the retention time of the cells within a single DRAM chip vary
significantly, resulting in \emph{strong cells} that have high retention
times and \emph{weak cells} that have low retention times within each chip.
The operating temperature affects the rate at which charge leaks from the
capacitor.  As the operating temperature increases, the retention time of a
DRAM cell decreases exponentially~\cite{liu.isca13, hamamoto.ted98}.}

Due to the rapid charge leakage from DRAM
cells, a DRAM controller periodically refreshes all DRAM cells
in place~\cite{liu.isca12, chang.hpca14, liu.isca13, jesd79.jedec13, qureshi.dsn15, khan.sigmetrics14, patel.isca17} (similar to
the periodic refresh techniques used in NAND flash memory, but at a much smaller
time scale). DRAM standards require a DRAM cell to be
refreshed once every \SI{64}{\milli\second}~\cite{jesd79.jedec13}. As the density of DRAM continues
to increase over successive product generations (e.g., by
128x between 1999 and 2017~\cite{chang.sigmetrics16, chang.thesis17}),
\chI{enabled by the scaling of DRAM to smaller manufacturing process
technology nodes\chVIII{~\cite{mandelman.ibmjrd02, mutlu.superfi14,mutlu.imw13, mutlu.date17}},} the performance
and energy overheads required to refresh an entire DRAM
module have grown significantly\chVIII{~\cite{mutlu.date17, liu.isca12, chang.hpca14, mutlu.superfi14,mutlu.imw13}}.
%  \chI{as DRAM scales to smaller
% manufacturing process technology nodes}.
\chI{It is expected that the refresh problem will get \chVIII{significantly} worse and \chI{limit}
DRAM density scaling, \chI{as described in a recent work by
Samsung and Intel~\cite{kang.mf14} and by our group~\cite{liu.isca12}}.
Prior analysis shows that when DRAM chip density reaches \SI{64}{\giga\bit},
nearly 50\% of the \chI{data} throughput is lost due to the high amount of time spent on
refreshing all of the rows in the chip, and
nearly 50\% of the DRAM chip power is spent on refresh operations~\cite{liu.isca12}.
Thus, \chVIII{data retention problems and refresh pose} a clear challenge to DRAM scalability.}

\chI{Various experimental studies of real DRAM chips (e.g., \chI{\cite{liu.isca12, liu.isca13, lee.hpca15, qureshi.dsn15, khan.sigmetrics14, khan.dsn16, patel.isca17, kim.iedl09, hassan.hpca17}})
have studied the \chI{data} retention time \chI{of DRAM cells in modern chips},
and have shown that the vast majority of DRAM cells can retain
data without loss for much longer than the \SI{64}{\milli\second} retention
time specified by DRAM standards.}
A number of works take advantage of this \chI{variability in data retention time behavior
across DRAM cells,
by reducing} the frequency at which the vast majority of DRAM \chI{rows} within a
module are refreshed \chI{(e.g., \chI{\cite{liu.isca12, liu.isca13, qureshi.dsn15, khan.sigmetrics14, venkatesan.hpca06, isen.micro09, patel.isca17, khan.cal16, baek.tc14}})}, or
by reducing the interference caused by
refresh requests on demand requests (e.g.,~\cite{chang.hpca14, mukundan.isca13, stuecheli.micro10}). 

More findings on the
nature of DRAM data retention and associated errors, as
well as relevant experimental data from modern DRAM
chips, can be found in our prior works\chI{~\cite{liu.isca12, chang.hpca14, liu.isca13, lee.hpca15, qureshi.dsn15, khan.sigmetrics14, khan.dsn16, patel.isca17, hassan.hpca17, chang.thesis17, khan.micro17, khan.cal16, mutlu.date17}}.
\chVIII{We also refer the readers to prior works on the design and operation of
the underlying DRAM architecture~\cite{kim.hpca10, kim.micro10, kim.isca12, lee.hpca13, lee.hpca15,
hassan.hpca16, liu.isca12, chang.hpca14, chang.hpca16, seshadri.micro13, chang.sigmetrics17,
lee.sigmetrics17, seshadri.micro17, liu.isca13, chang.sigmetrics16, hassan.hpca17, kim.cal15,
lee.pact15, lee.taco16, kim.isca14, patel.isca17, kim.hpca18}.}

% \sloppypar
\subsection{Errors in Emerging Nonvolatile Memory Technologies}
\label{sec:othermem:emerging}

DRAM operations are several orders of magnitude faster than
SSD operations, but DRAM has two major disadvantages. First,
DRAM offers orders of magnitude less storage density than
NAND-flash-memory-based SSDs. Second, DRAM is volatile
(i.e., the stored data is lost on a power outage). Emerging
nonvolatile memories, such as \emph{phase-change memory}
(PCM)~\cite{lee.isca09, qureshi.isca09, wong.procieee10, lee.micro10,
zhou.isca09, lee.comm10, yoon.taco14, meza.tr12},
\emph{spin-transfer torque
magnetic RAM} (STT-RAM or STT-MRAM)~\cite{naeimi.itj13, kultursay.ispass13}, \emph{metal-oxide
resistive RAM} (RRAM)~\cite{wong.procieee12}, and \emph{memristors}~\cite{chua.tct71, strukov.nature08},
are expected to bridge the gap between DRAM and SSDs, providing
DRAM-like access latency and energy, and at the same
time SSD-like large capacity and nonvolatility (and hence SSD-like
data persistence). 
\chI{These technologies are also expected to be used as part of
\emph{hybrid memory systems} \chI{(also called \emph{heterogeneous
memory systems})}, where one part of the memory consists of
DRAM modules and another part consists of modules of emerging
technologies\chI{~\cite{qureshi.isca09, yoon.taco14, ramos.ics11, yoon.iccd12, qureshi.micro12, 
zhang.pact09, meza.cal12, li.cluster17, meza.weed13, yu.micro17, chou.isca15, 
chou.micro14,jiang.hpca10, phadke.date11, chatterjee.micro12}}.}

PCM-based devices are expected to have
a limited lifetime, as PCM can only endure a certain number
of writes~\cite{lee.isca09, qureshi.isca09, wong.procieee10}, similar to the P/E cycling errors in
NAND-flash-memory-based SSDs (though PCM's write endurance
is higher than that of SSDs). PCM suffers from \chI{(1)}~\chI{\emph{resistance
drift}~\cite{wong.procieee10, pirovano.ted04, ielmini.ted07}}, where the resistance used to represent the value
\chI{becomes} higher over time (and eventually \chI{can introduce} a bit error),
similar to how charge leakage in NAND flash memory and
DRAM lead to retention errors over time; \chI{and
(2)~\chI{\emph{write disturb}~\cite{jiang.dsn14}}, where the heat generated during the programming of one
PCM cell dissipates into neighboring cells and \chI{can change} the value that is
stored within the neighboring cells}. 
\chI{STT-RAM suffers} from \chI{(1)}~\chI{\emph{retention failures}}, where the value
stored for a single bit \chI{(as the magnetic orientation of the layer that 
stores the bit)} can flip over time; and \chI{(2)}~\chI{\emph{read disturb}} (\chI{a
conceptually different phenomenon}
from the read disturb in DRAM and flash memory), where
reading a bit in STT-RAM can inadvertently induce a write to
that same bit~\cite{naeimi.itj13}. 

Due to the nascent nature of emerging
nonvolatile memory technologies and the lack of availability of
large-capacity devices built with them, extensive and dependable
experimental studies have yet to be conducted on the reliability
of real PCM, STT-RAM, RRAM, and memristor chips.
However, we believe that \chI{error mechanisms conceptually \chI{or
abstractly} similar} to those we
discussed for flash memory and DRAM are likely
to be prevalent in emerging technologies as well
\chI{(as supported by some recent studies~\cite{naeimi.itj13, jiang.dsn14, 
zhang.iccd12, khwa.isscc16, athmanathan.jetcas16, sills.vlsic15, sills.vlsit14})}, 
albeit with different underlying mechanisms and error rates.
\chVIII{We expect that the ROR and RFR techniques we propose in our HPCA 2015 paper~\cite{cai.hpca15} can
be easily adapted to NVM technologies.}

%!TEX root=paper.tex

\section{Significance}

Our HPCA 2015 paper~\cite{cai.hpca15} provides extensive characterization data
and proposes novel mechanisms to mitigate retention errors \chVIII{in modern NAND flash memory} and recover data
when ECC fails.  We believe that our characterization and mechanisms will have
a significant impact on the community, \chVIII{as evidenced by multiple recent
works directly building upon our HPCA 2015 paper\chIX{~\cite{cai.dsn15, jeong2015flashdefibrillator, luo.hpca18}}}.

%%------------------------------------------------------------------------------

% \subsection{Novelty}

% Our HPCA paper is the first to experimentally characterize and
% comprehensively analyze the distortion of threshold voltage distribution over
% different retention ages for state-of-the-art 2Y-nm MLC NAND flash memories.
% While other works have characterized data retention~\cite{cai.date12,
% mielke.irps08, papa.glsvlsi2014}, none of them have analyzed the results under
% room temperature at the comprehensive depth that we have.

% We are the first to propose an online technique, ROR, that dynamically optimizes
% the read reference voltage of each flash block based on its retention age. We
% are also the first to demonstrate insight into the flash read latency using a
% read-retry algorithm, and are the first to leverage these insights to reduce flash read latency.

% We are also the first to propose a technique, RFR, that \emph{enables} recovery from
% data loss resulting from a retention failure.  We are the first to characterize
% the difference between the leakage rate properties of flash cells, and we leverage
% this property in a novel way to recover damaged data after a retention failure happens.

\subsection{Long-Term Impact}

We believe our work will have long-term impact for the following
three reasons. First, as NAND flash memory becomes \chVIII{denser} in the future, data
retention will become a bigger issue, and thus a better understanding of its
implication and characteristics will be important to help maintain NAND flash reliability
after scaling\chVIII{~\cite{mutlu.date17,cai.procieee17,cai.procieee.arxiv17,cai.bookchapter.arxiv17}}. Second, we propose an online
technique \chVIII{that reduces} flash read latency, and we give insights into
the flash read-retry algorithm, thereby hopefully inspiring future works to further
optimize flash read latency. Third, we propose an offline technique that
leverages underlying flash characteristics to enable recovery from a retention
failure \chVIII{even after the drive fails to correct it}, thereby hopefully inspiring future works to look for more ways to
prevent data loss.

{\bf Data Retention.}  Our work provides a comprehensive analysis of the
retention loss effect on real NAND flash memory chips, which
enhances the understanding of the retention loss effect in the research
community. We hope that our analysis and solutions can inspire more works to
handle data retention in better ways. As planar NAND flash memory becomes denser, each
flash memory cell holds less charge and becomes more vulnerable to retention
loss~\cite{cai.date12, cai.iccd12}. Thus, in the future, we expect data retention
to become a more important problem\chVIII{~\cite{mutlu.date17,cai.procieee17,
cai.procieee.arxiv17, cai.bookchapter.arxiv17}}, and expect that industry will be more open to adapt
new solutions like our proposals, ROR and RFR. In fact, several flash-based SSDs
currently use refresh as a solution to mitigate retention
errors~\cite{techspot,hippo,muo}.
% \chVIII{and using refresh is recommended
% practice in SNIA~\cite{} \todo[I could not find this ref].}
\chVIII{Our work shows that
we can go significantly beyond refresh to tolerate the data retention problem in
NAND flash memory.}

{\bf Read Performance Optimization.} The read performance advantage of flash
memory over hard disk drives makes flash-based SSDs more appealing than hard disk
drives. However, many existing solutions, such as read-retry~\cite{cai.date13, fukami2017improving}, trade off flash
performance for reliability. Our HPCA 2015 paper~\cite{cai.hpca15} is the first to point
out the read performance problem, and to provide a detailed analysis and \chVIII{new} solution to
this problem. We hope that our work can enhance the research community's understanding of flash read
performance and bring more attention to flash read performance, \chVIII{which is
critically important to overall system performance. Techniques that are developed
in DRAM to reduce read latency~\cite{kim.hpca10, kim.micro10, kim.isca12, lee.hpca13, lee.hpca15,
hassan.hpca16, liu.isca12, chang.hpca14, chang.hpca16, seshadri.micro13, chang.sigmetrics17,
lee.sigmetrics17, seshadri.micro17, chang.sigmetrics16, 
lee.pact15, lee.taco16, kim.isca14, patel.isca17} can prompt inspiration for NAND flash memory.}

{\bf Data Recovery.} Prior to our work, after a retention failure happens,
\chVIII{an uncorrectable error and resulting data corruption} was considered
to be unrecoverable \chVIII{from}, resulting in data loss. To our knowledge, our HPCA 2015 paper~\cite{cai.hpca15} is the first to
show that it is actually \chVIII{\emph{possible}} to recover this data using our RFR mechanism.
As the reliability of NAND flash memory decreases, and the popularity of flash-based
SSDs increases, SSD failures are expected to increase, creating a greater need
for recovery techniques that can retrieve previously-unrecoverable data.
In light of this, recent works~\cite{cai.dsn15, jeong2015flashdefibrillator} have
\chVIII{directly} built upon RFR
to provide additional data recovery mechanisms.
We hope that our work draws more attention to flash memory data recovery, and
inspires further solutions to this important problem.

% \subsection{Related Recent Work}

% Our HPCA paper has inspired many recent work to (1)~characterize and study
% NAND flash memory errors, (2)~design new techniques to mitigate data retention
% errors in NAND flash memory, (3)~improve the optimal read reference
% voltage tuning technique, (4)~design new error recovery techniques for
% NAND flash memory, and etc.

\subsection{New Research Directions}

Our HPCA 2015 paper~\cite{cai.hpca15} presents characterization results for data retention in real NAND
flash chips. By making such data and knowledge available, we believe that the
flash \chVIII{memory and SSD} research \chVIII{communities} can have a better understanding of data retention, and
can therefore develop better solutions to tackle the retention problem in the future. We hope that our work will
continue to inspire future works in flash \chVIII{memory} that can provide a comprehensive
characterization and analysis of other NAND flash memory behavior using real
chips, such as program/erase cycling and cell-to-cell program disturbance.
We also hope that our ROR and RFR techniques bring more attention
to both the flash read performance problem and data recovery problem, and that they will inspire
researchers from both academia and industry to develop \chVIII{and adopt} new solutions. 

\section{Conclusion}

Our HPCA 2015 paper~\cite{cai.hpca15} comprehensively characterizes and analyzes how the
threshold voltage distribution and the optimal read reference
voltages of state-of-the-art 2Y-nm MLC NAND flash memory
change over different retention ages. Based on these analyses,
the paper proposes two new techniques. Retention Optimized
Reading (ROR) improves reliability, lifetime, and performance
of MLC NAND flash memory at modest storage cost by optimizing
the read reference voltage of each flash memory block
based on its retention age. We demonstrate significant benefits
with ROR in terms of reduced RBER, extended flash lifetime,
and reduction in flash read latency. Retention Failure Recovery
(RFR) recovers data with uncorrectable errors by identifying
and probabilistically correcting flash cells with retention errors.
We demonstrate large raw bit error rate reductions with RFR.
We hope that our comprehensive characterization of
data retention in flash memory will enable better understanding
of flash retention errors and motivate other new techniques to
overcome these errors. We believe the importance of our two
new techniques (ROR and RFR) will grow as NAND flash
memory scales to smaller feature sizes and becomes even less
reliable in the future.

\section*{Acknowledgments}

We \chVIII{thank Nandita Vijaykumar and the} anonymous reviewers for
feedback. This work is partially supported by the Intel Science
and Technology Center, CMU Data Storage Systems Center,
and NSF grants 1212962 and 1320531.

%%%%%%%%%%%%%%%%%%%%%%%%%%%%%%%%%%%%%%%%%%%%%%%%%%%%%%%%%%%%%%%%%%%%%%%%%%%%%%%%
% bibliography
%%%%%%%%%%%%%%%%%%%%%%%%%%%%%%%%%%%%%%%%%%%%%%%%%%%%%%%%%%%%%%%%%%%%%%%%%%%%%%%%
%\clearpage
% \setstretch{0.86}

\bibliographystyle{IEEEtranS}
\bibliography{ref}

\end{document}